\documentclass[12pt]{article}
\usepackage{amsmath}
\usepackage{amsfonts}
\usepackage{mathptm}
\usepackage{epsfig}
\usepackage{pst-plot,epsf}
\newcommand{\beq}{\begin{equation}}
\newcommand{\eeq}{\end{equation}}
\newcommand{\bea}{\begin{eqnarray}}
\newcommand{\eea}{\end{eqnarray}}
\newcommand{\nn}{\nonumber}

\newcommand{\epm}{e^+e^-}

\begin{document}
\thispagestyle{empty}
\begin{flushright}
%TP-USl-01/00\\
April 2015\\
Revised:\\
June 2015\\
\vspace*{1.cm}
\end{flushright}
\begin{center}
{\LARGE\bf {\tt carlomat\_3.0,}\\[4mm]
an automatic tool for the  electron--positron annihilation into
hadrons at low energies}\\
\vspace*{2cm}
%----------------------------
Karol Ko\l odziej\footnote{E-mail: karol.kolodziej@us.edu.pl}\\[1cm]
{\small\it
Institute of Physics, University of Silesia\\ 
ul. Uniwersytecka 4, PL-40007 Katowice, Poland}\\
\vspace*{3.cm}
{\bf Abstract}\\
\end{center}
A new version of {\tt carlomat} that allows to generate automatically
the Monte Carlo programs dedicated to the description of the processes 
$\epm\;\to\;{\rm hadrons}$ at low center-of-mass energies is presented. 
The program has been substantially modified in order to incorporate the
photon--vector meson mixing terms and to make possible computation of 
the helicity amplitudes involving the Feynman interaction vertices of new 
tensor structures, like those predicted by the Resonance Chiral Theory or 
Hidden Local Symmetry model, and the effective Lagrangian of the 
electromagnetic interaction of the nucleons. Moreover, a number of new options 
have been introduced in the program in order to enable
a better control over the effective models implemented. In particular,
they offer a possibility to determine the dominant production mechanisms of 
the final state chosen by the user.

\vfill
\newpage

{\large \bf PROGRAM SUMMARY}\\[4mm]
{\it Program title:} {\tt carlomat, version 3.0}\\
{\it Catalogue identifier:}\\
{\it Program summary URL:}\\
{\it Program obtainable from:} CPC Program Library, Queen's University, Belfast,
N. Ireland\\
{\it Licensing provisions:} Standard CPC licence\\
{\it No. of lines in distributed program, including test data, etc.:}\\
{\it No. of bytes in distributed program, including test data, etc.:}\\
{\it Distribution format:} tar.gz\\
{\it Programming language:} {\tt Fortran 90/95}\\
{\it Computer:} All\\
{\it Operating system:} Linux\\
{\it Classification:}\\
{\it Nature of problem:}\\ 
Predictions for reactions of low energy $\epm$-annihilation into final states 
containing pions, kaons, light vector mesons, one or more photons and light 
fermion pairs within the Standard Model and effective models inspired by 
the Resonance Chiral Theory or Hidden Local Symmetry model. 
Description of the electromagnetic production of nucleon pairs
within the effective Lagrangian approach.\\
{\it Solution method:}\\
As in former versions, a program for the Monte Carlo (MC) simulation
of $\epm\;\to\;{\rm hadrons}$ at low energies is
generated in a fully automatic way for a user specified process. However, 
the user is supposed to select a number of options and adjust arbitrary
parameters in the main part of the MC computation program in order to obtain 
possibly the best description of experimental data. To this end, the user can 
also easily supplement her/his own formulae for $s$-dependent vector meson 
widths or running couplings by appropriately modifying corresponding
subroutines.\\
{\it Reasons for new version:}\\
Processes of $\epm\;\to\;{\rm hadrons}$ in the energy range
below  the $J/\psi$ threshold cannot be described in the framework of
perturbative quantum chromodynamics. The scalar 
electrodynamics which has been implemented in {\tt carlomat\_2.0} 
\cite{carlomat2} does not provide a satisfactory description either. 
The most promising theoretical frameworks in this context are
the Resonance Chiral Theory or Hidden Local
Symmetry model which, among others, involve the photon--vector meson mixing 
and a number of vertices of rather complicated Lorentz tensor structure that 
is not present in the Standard Model or scalar QED. Already at low energies, 
the hadronic final states may consist of several particles, such as pions, 
kaons, or nucleons which can be accompanied by one or more photons, or light 
fermion pairs such as $e^+e^-$, or $\mu^+\mu^-$. The number of Feynman diagrams
of such multiparticle reactions grows substantially with increasing numbers of 
interaction vertices and mixing terms of the effective models. Therefore, it 
is highly desirable to automatize the calculations. At the same time,
new program options should provide the user with an easy way of implementing
her/his own changes in the program in order to better fit the experimental 
data. \\
{\it Summary of revisions:}\\
The code-generation part of the program has been substantially modified 
in order to incorporate the photon--vector meson mixing and 
calls to new subroutines for computation of the helicity amplitudes of
the building blocks and complete Feynman diagrams which contain new
interaction vertices and mixing terms. The subroutine library of {\tt carlomat}
has been extended to make possible computation of the helicity amplitudes 
involving the Feynman interaction vertices of new Lorentz tensor structures.
Many subroutines have been modified in order to incorporate the $q^2$-dependent
couplings and vector meson widths. A number of options have been introduced 
in order to give a better control of the effective model implemented.\\
{\it Restrictions:}\\
As in previous versions of the program the number of particles is limited to 12
which exceeds typical numbers of
particles of the exclusive low energy $\epm$-annihilation processes.
However, in the presence of photon--vector meson mixing, the 
Feynman diagrams proliferate, for example, with currently implemented
Feynman rules, there are 90672 diagrams of $\epm\;\to\;3(\pi^+\pi^-)$. 
Hence, the compilation time of generated
code may become very long already for processes with smaller 
number of the final state particles.
Many couplings of the effective models are not known with good enough 
precision and must be adjusted in consecutive runs of the program in order 
to obtain satisfactory description of the experimental data.\\
{\it Running time:}\\ 
Depends on the selected process. Typical running time for the code generation
vary from a fraction of a second for, e.g., $\epm\;\to\;\pi^+\pi^-K^+K^-$
to about 2 minutes for $\epm\;\to\;3(\pi^+\pi^-)$. It may become substantially
longer for processes with more particles in the final state.
The execution time necessary to produce the appended test output files
for $e^+ e^- \to \pi^+ \pi^- \mu^+ \mu^- \gamma$ and 
$e^+ e^- \to \pi^+ \pi^- \pi^+ \pi^- \gamma$ was 13s and 4s, respectively.
The code generation for both processes took a fraction of a second time
for each process.
\section{Introduction}

Hadronic contributions to the vacuum polarization are the major factor that 
influences precision of theoretical predictions for the muon anomaly 
$a_{\mu}$ and plays an important role in the evolution of 
the fine structure constant $\alpha(Q^2)$ from the Thomson limit to high energy 
scales. Improving the precision of predictions for the muon anomaly 
becomes vital in the prospect of forthcoming measurements in 
Fermilab that should reduce the experimental error of $a_{\mu}$ to 0.14 parts 
per million, while the better precision of $\alpha(m_Z^2)$ would be important 
for the precision data analysis from the future high energy $\epm$ collider, 
which would most probably include a giga-$Z$ option.
%the project being more and more seriously discussed in the international 
%community of high energy physicists. 
Because of the breakdown of predictive power of the perturbative QCD at low 
momentum transfer, the hadronic contributions to the vacuum polarization are 
determined, with the help of dispersion relations, from the energy dependence 
of the total 
cross section of electron--positron annihilation into hadrons, 
$\sigma_{e^+e^-\to {\rm hadrons}}(s)$. Below  the $J/\psi$ production threshold, 
$\sigma_{e^+e^-\to {\rm hadrons}}$ must be measured and confronted with 
theoretical predictions of some effective model for the low energy hadron 
physics.

There are two QCD inspired theoretical frameworks which seem to 
be applicable in this context: 
the Resonance Chiral Theory (R$\chi$T) \cite {RChPT} and 
the Hidden Local Symmetry (HLS) model \cite{HLS}, which were proven to be 
essentially equivalent \cite{equiv}.  For example, the HLS model allowed for 
a quite satisfactory simultaneous description of most hadronic
$\epm$-annihilation channels in the low energy range, including $\phi$-resonance
and 10 decay widths, mostly radiative ones, of light mesons and allowed 
to resolve the inconsistency between the $\epm$-annihilation to $\pi^+\pi^-$ 
and the $\tau^{\pm}$-decay to $\pi^{\pm}\pi^0\nu_{\tau}$ \cite{Benayoun1}, 
\cite{Benayoun2}. The hadronic currents based on R$\chi$T were 
implemented in {\tt TAUOLA}, a $\tau$-decay Monte Carlo (MC) 
generator \cite{TAUOLA}, 
and used for description of the $\tau$ lepton decay into two or three 
pseudoscalar mesons that constitute 88\% of the $\tau$ hadronic decay width 
in Ref.~\cite{taudec1} and later improved for 
$\tau^{\pm}\;\to\;\pi^{\pm}\pi^{\pm}\pi^{\mp}\nu_{\tau}$ decay mode
in Ref.~\cite{taudec2} which allowed to successfully model the one-dimensional 
distributions measured by the BaBar collaboration.

The number of Feynman diagrams in the framework of R$\chi$T or HLS model grows 
quite fast with the number of particles in the final state of 
$\epm\;\to\;{\rm hadrons}$.
In particular, in the presence of one or a few photon--vector meson mixing 
terms, it can easily reach a hundred thousand already for $\epm\;\to\;6\pi$. 
Obviously, preparation of a reliable MC generator for such a process is 
rather tedious a task, 
unless the process of code writing is fully automatized. The first step toward 
the automatic generation of the MC programs for 
description of reactions $\epm\;\to\;{\rm hadrons}$ 
was already made in {\tt carlomat\_2.0} \cite{carlomat2}, 
in which the Feynman rules of the scalar electrodynamics (sQED) were implemented
in addition to those of the Standard Model (SM).
It allowed to effectively describe the electromagnetic (EM) interaction
of charged pions which, at low energies, can be considered as being point like 
particles, see, e.g., \cite{fred}. However, the charged pion 
form factor that would allow to account for the bound state nature of the 
particle was not implemented in {\tt carlomat\_2.0}. In the present paper, 
a new version, labeled with 3.0, of a program {\tt carlomat} \cite{carlomat}, 
\cite{carlomat2} is described, which to large extent 
should meet the requirements of automatic code generation for MC simulation
of the low energetic $\epm$-annihilation into hadrons in the framework of 
the effective models.

The paper is organized in the following way. In Section 2, the implementation 
of new Feynman rules in the program is described.
New options which have been implemented in the program to give 
the user a better control over the model are described in Section 3. Finally,
the instructions for preparation for running and usage of the
program are given in Section 4.

\section{New Feynman rules implemented in the program}
In this section, the implementation of the Feynman rules of the HLS model that 
are relevant for the description of $e^+e^-\to {\rm hadrons}$ in the low 
energy range in {\tt carlomat\_3.0} is described.
Most of the rules can be derived from the Lagrangian pieces of Appendix C of 
Ref.~\cite{Benayoun1}. The Lagrangian of EM interaction 
of spin 1/2 nucleons implemented in the program is described in Subsection 2.3.

\subsection{Photon--vector meson mixing}
\begin{figure}[!ht]
\centerline{
\epsfig{file=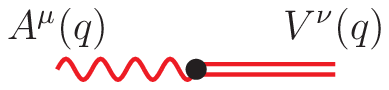,  width=45mm, height=9mm}
$\quad\equiv\quad -e f_{AV}(q^2)\;g^{\mu\nu},\quad$ with 
$\quad V=\rho^0, \omega, \phi, \rho_1, \rho_2.$}
\caption{\small The photon--vector meson mixing diagrams implemented in the current 
version of the program;  $\rho_1$ and $\rho_2$ stand for $\rho(1450)$ and
$\rho(1700)$, respectively.}
\label{mixing}
\end{figure}

The topology generator of {\tt carlomat} takes into account only triple 
and quartic vertices, therefore the mixing should be added in subroutine 
{\tt checktop}, where topologies of diagrams are confronted with the 
implemented Feynman rules. The procedure was described in detail in 
Ref.~\cite{carlomat}. For the sake of clarity 
let us remind here, that every topology in {\tt carlomat} is divided into two 
parts, each being checked against the Feynman rules separately. This is done by 
consecutive calls to subroutine {\tt genpart} that
combines two (three) particles into the third (fourth)
leg of a triple (quartic) Feynman vertex which is then folded with 
the adjacent Feynman propagator to form an off-shell particle.
The latter is represented by an array of spinors, polarization vectors 
or scalars, 
whose elements are labeled with different combinations of the polarization
indices of the particle spinors or polarization vectors of which they are 
formed. At this point, if the particle mixing is present, a new subroutine 
{\tt mixpl} is called to check whether the propagator
of the off-shell particle can be mixed with some other propagator or not, 
if so, a new off-shell particle is formed. The particle with mixing is 
appropriately tagged in order not to be mixed again, because, according to 
Fig.~\ref{mixing}, the mixing term contains an extra power of the electric 
charge $e$ and therefore should be considered as the next to leading-order 
correction. This procedure is being
repeated until finally two off-shell particles are formed, corresponding
to both parts of the considered topology. Then, a modified subroutine 
{\tt matchkk} is called which checks, whether the particles can be matched 
in the Feynman diagram with the Feynman propagator or, if none of them
has been mixed yet, with some of the mixing terms of Fig.~\ref{mixing}.

In spite of being conceptually quite simple, the implementation of particle mixing 
required substantial changes in the code-generation part of the program.
Moreover, new subroutines {\tt bbkk} and {\tt bbmd} have been written
to compute, respectively, the polarization vectors of the off-shell particle
and helicity amplitudes of the Feynman diagrams in the case of mixing.

\subsection{Interaction vertices}
\begin{figure}[!ht]
\centerline{
\epsfig{file=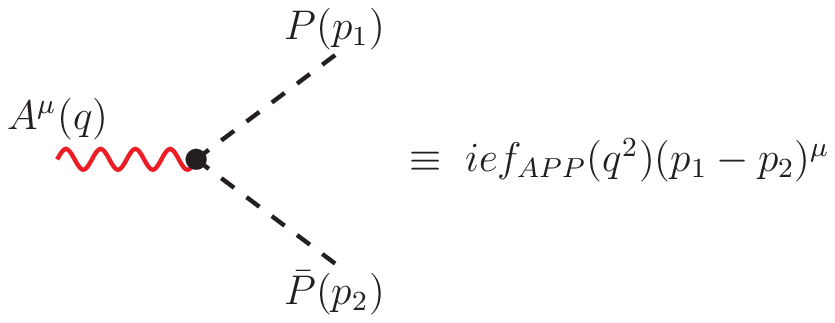,  width=75mm, height=30mm} \hspace*{1cm}
\epsfig{file=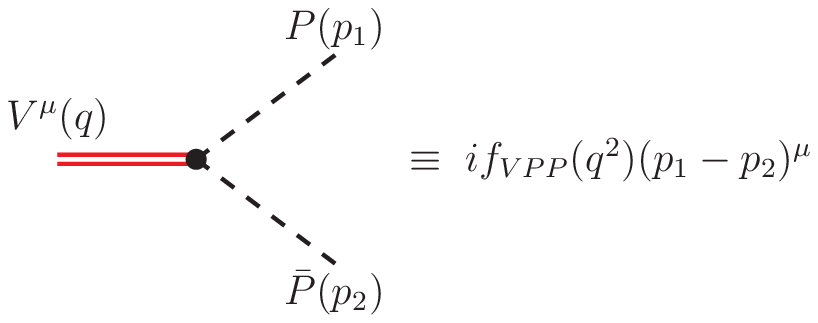,  width=75mm, height=30mm} 
}
\caption{\small Triple vertices of the photon $A^{\mu}$ or vector meson 
$V^{\mu}$, $V=\rho^0, \omega, \phi$, interaction with the pseudoscalar meson 
pair $P\bar P$, $P=\pi^+,K^+,K^0$, of the same form as that of the triple vertex
of sQED.}
\label{vertsQED}
\end{figure}

\begin{figure}[!ht]
\centerline{
\epsfig{file=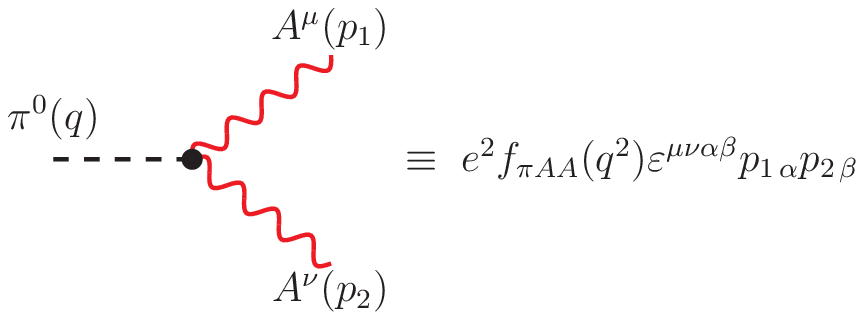,  width=75mm, height=30mm} \hspace*{1cm}
\epsfig{file=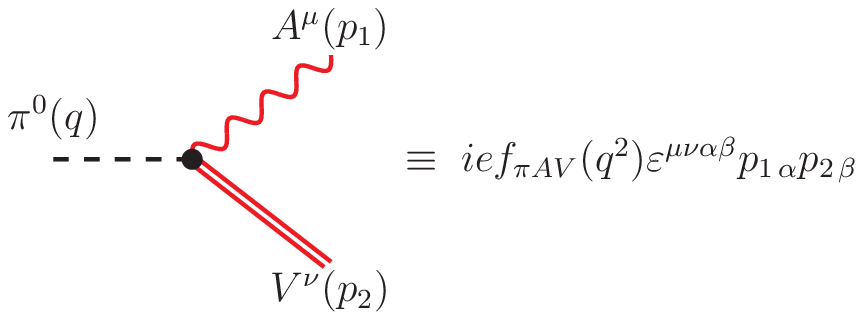,  width=75mm, height=30mm}}  
\vspace*{1cm}
\centerline{
\epsfig{file=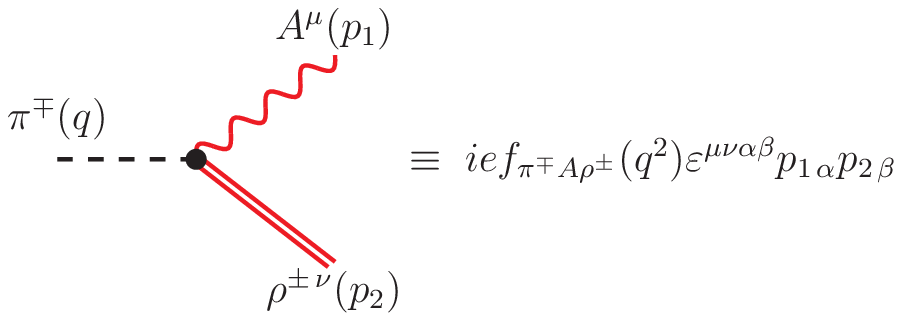,  width=75mm, height=30mm} \hspace*{1cm}
\epsfig{file=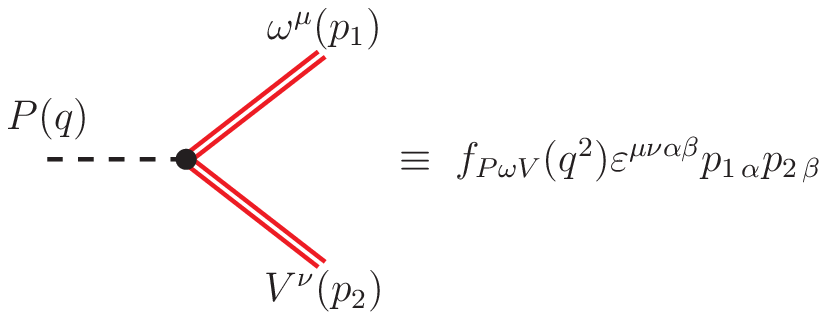,  width=75mm, height=30mm}  
}
\caption{\small Triple vertices of the pion interaction with photons
or vector mesons, where, in the 
top right corner, $V=\rho^0,\omega$, in the bottom right corner $P=\pi^0$ 
and $V=\rho^0$ or $P=\pi^{\mp}$ and $V=\rho^{\pm}$.}
\label{verts3}
\end{figure}

\begin{figure}[!ht]
\begin{tabular}{cc}
%\centerline{
\epsfig{file=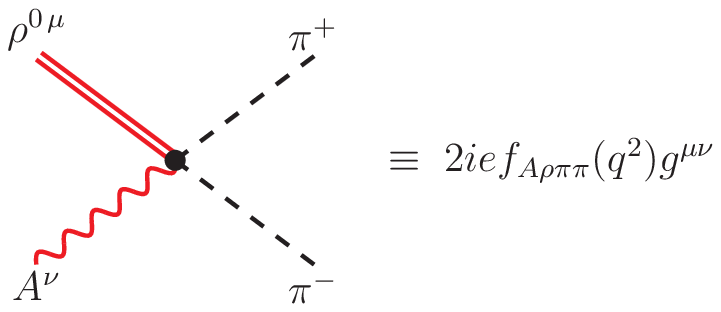,  width=65mm, height=30mm} &%\hspace*{2cm}
\epsfig{file=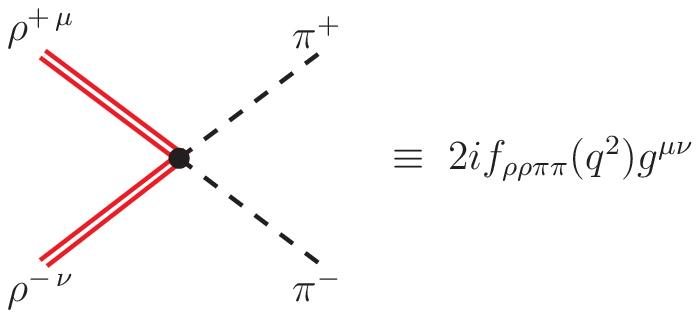,  width=65mm, height=30mm}\\[5mm]%}  
%\vspace*{1cm}
\hspace*{0.4cm}
\epsfig{file=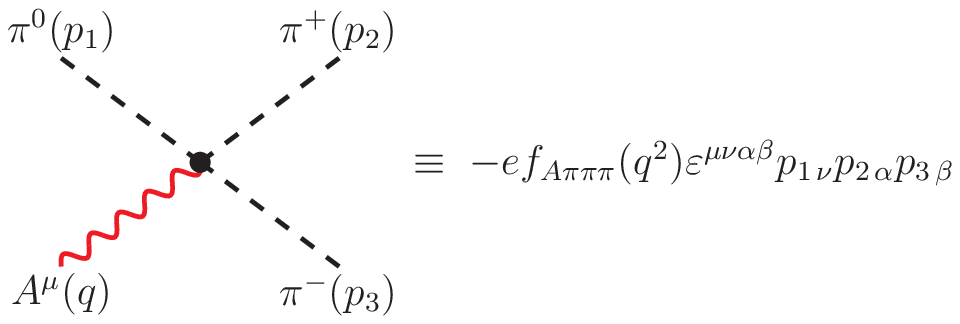,  width=75mm, height=30mm} &
\hspace*{0.4cm}
\epsfig{file=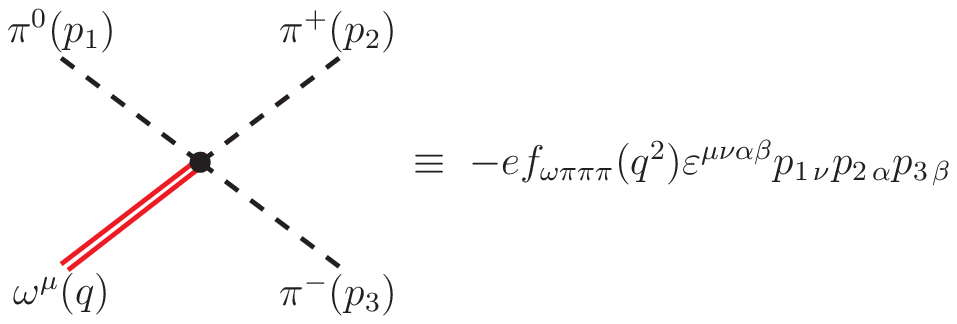,  width=75mm, height=30mm}  
\end{tabular}
\caption{\small Quartic vertices of the HLS model implemented in the current
version of the program. The quartic vertex $AA\pi^+\pi^-$ of sQED, implemented 
already in {\tt carlomat\_2.0}, is not shown.}
\label{verts4}
\end{figure}

The triple and quartic interaction vertices of the %R$\chi$T or
HLS model that are implemented in {\tt carlomat\_3.0} are depicted 
in Figs.~\ref{vertsQED}--\ref{verts4}, where all the particle four momenta 
are assumed to be incoming to the vertex and $\varepsilon^{\mu\nu\rho\sigma}$,
with $\varepsilon^{0123}=1$, is the totally antisymmetric Levi-Civita tensor. 
A number of new subroutines for computation of the building blocks and complete 
amplitudes of the Feynman diagrams containing vertices of new tensor structure 
have been written. The implementation of calls to the new subroutines
required some changes in the code-generation part of the program which
concerned mainly subroutine {\tt genpart}.

The triple vertices of the photon $A^{\mu}$ or vector meson 
$V^{\mu}$ interaction with the pseudoscalar meson pair $P\bar P$ which have the 
form similar to the triple vertex of sQED are shown in Fig.~\ref{vertsQED}. The
only difference is the replacement
\bea
\label{replacement}
e\;\to\;ef_{APP}(q^2)\qquad {\rm and}\qquad e\;\to\;f_{VPP}(q^2),
\eea
where $V=\rho^0, \omega, \phi, \rho_1, \rho_2$ and $P=\pi^+,K^+,K^0$. 
Although couplings of $\rho_1=\rho(1450)$ and $\rho_2=\rho(1700)$ to other 
particles are hard to define on the basis of existing data \cite{PDG}, the 
interaction vertices $\rho_i\pi^+\pi^-$ and mixing terms $\gamma-\rho_i$, 
$i=1,2$, have been included in the program just 
to enable tests of their possible influence on some observables, e.g., on 
the pion form factor, where they play a role. 

Subroutines {\tt ppakk, appkk}
and {\tt papkk} for the computation of building blocks of the Feynman diagrams, 
and {\tt ppamd, appmd} and {\tt papmd} for the computation of the helicity 
amplitudes in the sQED of {\tt carlomat\_2.0} have been all supplemented with an
option {\tt ig}, which allows to take into account the $q^2$-dependent couplings
of (\ref{replacement}). Subroutines {\tt ppakk} and {\tt ppamd} have been 
additionally supplied with an option {\tt iwdth}, that gives a possibility to 
include the $s$-dependent width of a vector meson. The use of both options 
is explained in Section 3.

Triple interaction vertices of the HLS model that have a form different from
that of the triple vertices of the SM or sQED are depicted
in Fig.~\ref{verts3}. New subroutines that have been written in order
to compute the corresponding building blocks and helicity amplitudes 
are: {\tt pvvkk}, {\tt pvvmd}, {\tt vvpkk} and {\tt vvpmd}. All of them
include the running-coupling option {\tt ig}, and the
first one, whose output is an array of four vectors corresponding to all 
possible helicities of the scalar and vector particles they are composed of,
includes in addition the running-width option {\tt iwdth}.

The quartic interaction vertices of the HLS model implemented in the current
version of the program are shown in Fig.~\ref{verts4}. 
The vertices in the first row have the same tensor form as
the quartic vertex of the sQED or the quartic vertices of the Nambu-Goldstone boson
-- gauge boson interaction of the SM, which have been implemented already
in the first version of {\tt carlomat}.
Hence, the corresponding building blocks and helicity amplitudes 
can be computed with modified subroutines {\tt vvsskk, vvssmd, vsvskk, vsvsmd,
vssvkk, vssvmd, svvskk, svvsmd, ssvvkk, ssvvmd, svsvkk} and {\tt svsvmd}, 
which have been all supplied with the running-coupling option {\tt ig}. Subroutines
{\tt vssvkk, ssvvkk} and {\tt svsvkk} have been moreover supplemented with
the running-width option {\tt iwidth}. 
The tensor form of the vertices in the second row of Fig.~\ref{verts4}
is different. Therefore, the corresponding building blocks and helicity 
amplitudes are computed with newly written subroutines {\tt pppvkk, pppvmd,
vpppkk} and {\tt vpppmd}.
\subsection{Electromagnetic interaction of nucleons}
The Lagrangian of EM interaction of spin 1/2 nucleons has the following
form:
\bea
\label{gammaNN}
{\mathcal L}_{ANN}&=&eA_{\mu}\bar{N}(p')\left[\gamma^{\mu} F_1(Q^2)
+\frac{i}{2m_N}\sigma^{\mu\nu}q_{\nu} F_2(Q^2)\right]N(p),
\eea
where $\sigma^{\mu\nu}=\frac{i}{2}[\gamma^{\mu},\gamma^{\nu}]$, with $\gamma^{\mu}$,
$\mu=0,1,2,3$, being the Dirac matrices, $q=p-p'$
is the four momentum transfer, $F_1(Q^2)$ and $F_2(Q^2)$ are the form factors 
and $Q^2=-q^2$. The form of Eq.~(\ref{gammaNN}) is similar 
to that of the effective Lagrangian of the $Wtb$ interaction given
by Eq.~(4) of Ref.~\cite{carlomat2}.
Due to this fact, the implementation of the corresponding Feynman rules
for the nucleon--photon interaction was straightforward. 
To compute the corresponding building blocks and helicity amplitudes 
the following new subroutines have been written: {\tt annkk, annmd, nnakk,
nnamd, nankk} and {\tt nanmd} and the calls to them have been appropriately
implemented in subroutine {\tt genpart}. The form factors $F_1(Q^2)$ and 
$F_2(Q^2)$ have been adopted from {\tt PHOKARA} \cite{PHOKARA} with the help
of an interface 
subroutine {\tt nuclff\_phok}. In this way, the MC simulations of processes 
involving the EM production of the nucleon pairs have become possible.
\section{New program options}
New options which have been added in the program to give the user a better 
control over the implemented models for the description of 
the electron--positron annihilation into hadrons at low energies are explained 
below.

All subroutines that are used to compute the building blocks or the complete
helicity amplitudes of the Feynman diagrams of 
Figs.~\ref{vertsQED}--\ref{verts4}
have been supplied with the running-coupling option the name of which
is formed by adding a prefix {\tt i} to the name of the corresponding coupling, 
as the name is created in exactly the same way
at the stage of code generation. The options are to be specified in subroutine
{\tt couplsm}, where they are defined below the assignment instruction for
each particular coupling.

\begin{tabular}{ll}
{\tt icoupl\_name=0/1,2,...} & if the {\tt fixed/running} coupling is to be 
                               used in the computation,
\end{tabular}

where choices {\tt 1,2,...} corresponding to different running couplings 
$f_{...}(q^2)$ of Figs.~\ref{mixing}--\ref{verts4} should be added by the user 
as extra {\tt else~if~(ig == ...)~then} blocks in subroutine {\tt runcoupl}.
The block must contain an assignment for a double complex variable {\tt rg} in 
terms
of the four momentum transfer squared $q^2$ and any other physical parameters
that are available in module {\tt inprms}.
The actual form of the four momentum transfer $q$ is determined automatically 
from the four momentum conservation in the corresponding interaction vertex at 
the stage of code generation. Many couplings of the R$\chi$T or HLS model are 
not known well enough and therefore
must be adjusted in consecutive runs of the program in order to obtain
satisfactory description of the experimental data. 
If there are no hints as to the form of the running couplings $f_{...}(q^2)$
then it is recommended to set the corresponding running-coupling option to 0,
which means that the fixed coupling is to be used in the computation.
The user can also modify any of the fixed couplings by changing the 
corresponding 
assignments in {\tt couplsm}, where the couplings are defined in terms of the 
physical parameters of module {\tt inprms}.

The subroutines for computation of the four vectors representing vector mesons 
have been in addition supplied with the running-width
option {\tt iwdth\_name}, i.e. {\tt igmrh, igmom, igmph, igmr1, igmr2} for the 
running width of $\rho^0,\omega,\phi,\rho_1,\rho_2$, respectively:

\begin{tabular}{ll}
{\tt iwdth\_name=0/1,2,3} & if the {\tt fixed/running} width of the vector 
                            particle should be used,
\end{tabular}

where choices 1,2,3 refer to different running-width options in subroutine 
{\tt runwidth} which again can easily be extended by the user.
The options are controlled from {\tt carlocom},
the main part of the MC computation program.

The main part of the MC computation program 
{\tt carlocom} contains a few flags:
{\tt iarho, iaome, iaphi, iarho1} and {\tt iarho2} that allow to switch off and 
on the photon mixing with $\rho, \omega, \phi,$ $\rho_1$ and $\rho_2$ vector 
mesons without a need of running the code-generation program anew,  
provided that the corresponding mixing terms were included in a file
{\tt vertices.dat} when the MC code was generated. This gives a possibility
to determine the dominant production mechanisms of the final state 
considered by the user.

In order to give a better control over the mixing contributions to a given 
process, subroutines {\tt bbkk} and {\tt bbmd} are equipped with the option:

\begin{tabular}{ll}
{\tt iwgt=0/1,2,...}  & if the additional complex factor $c_1,c_2,...$
                        {\tt is not/is} to be included in $f_{AV}(q^2)$\\ 
                      & of Fig.~\ref{mixing}, i.e. in the amplitude of
                        the Feynman diagrams containing this\\
                      & particular particle mixing contribution.
\end{tabular}

The actual names for that option in {\tt carlocom} are: 
{\tt imrho, imome, imphi, imrh1, imrh2} for the 
$\rho^0,\omega,\phi,\rho_1,\rho_2$ meson, respectively.
The complex factor $c_j$, $j=1,2,...$ is given by
\bea
c_j=w_j\;e^{i\varphi_j}f_j(q^2),
\eea
where $w_j$ is a positive weight, $\varphi_j$ is an angle in degrees, which should
be both specified for each possible particle mixing term in the main program for
the MC computation {\tt carlocom}, and $f_j(q^2)$ is a possible four momentum
transfer dependence that is defined in subroutine {\tt weightfactor}. Actually 
only three simple dependences corresponding to {\tt iwgt=1,2,3} are currently 
defined in {\tt weightfactor}, but the user can easily add more options by
implementing new {\tt else~if~(iwgt == ...)~then} conditions. 

An important new option in the program, which allows to test the EM 
gauge invariance for processes with one or more external photons, is
{\tt igauge} in {\tt carlocom.f}:

\begin{tabular}{ll}
{\tt igauge=1,2,.../else} & if the gauge invariance {\tt is/is not} to be 
tested, 
\end{tabular}

where 1,2,... is the number of a photon, counting from left to right,
whose polarization four vector is replaced with its four momentum.

\begin{figure}[!ht]
\centerline{
\epsfig{file=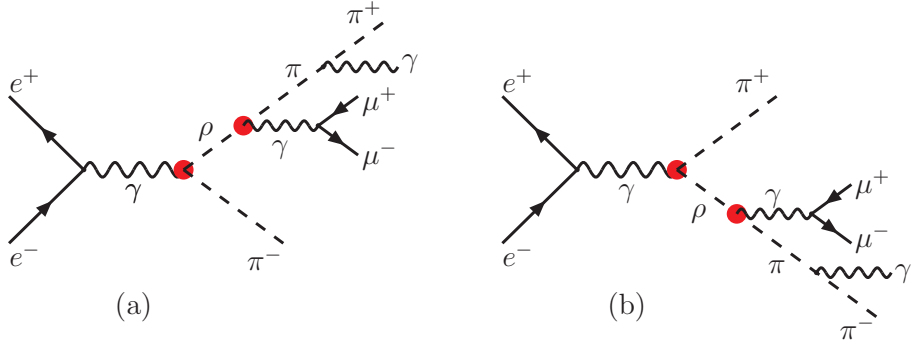,  width=120mm, height=45mm}}
\caption{\small The Feynman diagrams of process (\ref{epmppmmg}) that spoil
the EM gauge invariance. The blobs indicate the vertices $\pi^{\mp}\gamma\rho^{\pm}$.}
\label{diags}
\end{figure}

To illustrate how this option can be used in practice, consider the following 
radiative processes:
\bea
\label{epmppmmg}
e^+ e^- &\to& \pi^+ \pi^- \mu^+ \mu^- \gamma,\\
\label{epm4pg}
e^+ e^- &\to& \pi^+ \pi^- \pi^+ \pi^- \gamma.
\eea
Taking into account the Feynman rules of SM, without the Higgs couplings
to electrons and muons, sQED, the 
$\gamma - \rho^0$ mixing of Fig.~\ref{mixing} and the vertices:
$\gamma\pi^+\pi^-$ and $\rho^0\pi^+\pi^-$ of Fig.~\ref{vertsQED},
$\pi^0\gamma\gamma$ and $\pi^0\gamma\rho^0$ of Fig.~\ref{verts3} and
$\gamma\rho^0\pi^+\pi^-$ and $\gamma\pi^0\pi^+\pi^-$ of Fig.~\ref{verts4},
processes (\ref{epmppmmg}) and (\ref{epm4pg}) receive contribution from, 
respectively, 209 and 774 Feynman diagrams. If, in addition, the vertices 
$\pi^{\mp}\gamma\rho^{\pm}$ of Fig.~\ref{verts3} are included then the number 
of diagrams of processes (\ref{epmppmmg}) and (\ref{epm4pg}) grows, respectively, 
to 231 and 968. The cross sections of processes (\ref{epmppmmg}) 
and (\ref{epm4pg}) at $\sqrt{s}=1$~GeV, with the following cuts on the
angles between the photon and a lepton $\theta_{\gamma\,l}$, the photon 
and a pion $\theta_{\pi\,l}$ and the photon energy:
\bea
\label{cuts}
\theta_{\gamma\,l} > 5^{\circ}, \qquad \theta_{\gamma\,\pi} > 5^{\circ}, \qquad
E_{\gamma} > 10\;{\rm MeV},
\eea
are presented in Table~1. The cross sections without (with) 
contribution from the $\pi^{\mp}\gamma\rho^{\pm}$ interaction vertices
of Fig.~\ref{verts3} are printed in the first (second) column for each
process. If {\tt igauge=1} then the cross section
drops by about 32 orders of magnitude, which means that the EM gauge invariance
works perfectly well. However, if the vertices $\pi^{\mp}\gamma\rho^{\pm}$
of Fig.~\ref{verts3} are included then the EM gauge invariance is not so
perfect any more.
For process (\ref{epmppmmg}) this is caused by the two Feynman diagrams depicted
in Fig.~\ref{diags}. To justify this statement, let us denote the four
momenta of particles of process (\ref{epmppmmg}) by 
$p_1,p_2,...,p_7$, from left to right
consecutively, and consider the EM gauge invariance test for 
the amplitudes of the diagrams (a) and (b) of
Fig.~\ref{diags}, which means in practice that the photon polarization four vector 
is replaced with its four momentum. Neglecting the $i$ factors, which are the 
same for both amplitudes, and skipping polarization indices the amplitudes read:
\bea
\label{a}
M_a&=&g^2
\varepsilon_{12\,\nu}\,\varepsilon^{\nu\mu\alpha\beta}p_{12\,\alpha}
(-q_{\beta})\,\frac{-g_{\mu\rho}+\frac{q_{\mu}q_{\rho}}{M^2}}{q^2-M^2}\,
\varepsilon_{56\,\sigma}\,\varepsilon^{\sigma\rho\gamma\delta}(-p_{56\,\gamma})q_{\delta}
\,s_{37}\nn\\
&=&\frac{eg^2}{q^2-M^2}\,\varepsilon^{\mu\nu\alpha\beta}\,
\varepsilon_{\mu\sigma\gamma\delta}\,
\varepsilon_{12\,\nu}\,p_{12\,\alpha}\,p_{4\,\beta}\,
\varepsilon_{56}^{\sigma}\,p_{56}^{\gamma}(p_3+p_7)^{\delta},\\
\label{b}
M_b&=&g^2
\varepsilon_{12\,\nu}\,\varepsilon^{\nu\mu\alpha\beta}p_{12\,\alpha}
(-r_{\beta})\,\frac{-g_{\mu\rho}+\frac{r_{\mu}r_{\rho}}{M^2}}{r^2-M^2}\,
\varepsilon_{56\,\sigma}\,\varepsilon^{\sigma\rho\gamma\delta}(-p_{56\,\gamma})r_{\delta}
\,s_{47}\nn\\
&=&-\frac{eg^2}{r^2-M^2}\,\varepsilon^{\mu\nu\alpha\beta}\,
\varepsilon_{\mu\sigma\gamma\delta}\,
\varepsilon_{12\,\nu}\,p_{12\,\alpha}\,p_{3\,\beta}\,
\varepsilon_{56}^{\sigma}\,p_{56}^{\gamma}(p_4+p_7)^{\delta},
\eea
where $\varepsilon_{12}^{\nu}$ $(\varepsilon_{56}^{\sigma})$ is the polarization
four vector representing the $e^+e^-\gamma$ $(\mu^-\mu^+\gamma)$ vertex 
contracted with the adjacent photon propagator,
$M^2=m_{\rho}^2-i m_{\rho}\Gamma_{\rho}$ is the complex $\rho$ meson mass 
parameter, $p_{12}=p_1+p_2$, $p_{56}=p_5+p_6$, 
$q=p_{56}+p_3+p_7$ and $r=p_{56}+p_4+p_7$ are four momenta of intermediate
virtual photons and $\rho^{\pm}$ mesons, 
and the coupling $g=ef_{\pi^-A\rho^+}(q^2)=ef_{\pi^+A\rho^-}(q^2)
=ef_{\pi^-A\rho^+}(r^2)=ef_{\pi^+A\rho^-}(r^2)$ has been assumed to have a fixed
value. In the second row of Eqs.~(\ref{a}) and (\ref{b}), use has been made 
of the fact that,
in the EM gauge invariance test, the scalars $s_{37}$ and $s_{47}$ representing 
the $\pi^+\pi^-\gamma$ and $\pi^-\pi^+\gamma$ vertex multiplied with 
the adjacent pion propagator, take the following form:
\bea
s_{37}&=&e\,
\left.\frac{(p_3+p_7-(-p_3))^{\mu}\,\varepsilon_{\mu}^*(p_7)}{(p_3+p_7)^2-m_{\pi}^2}
\right|_{\varepsilon(p_7)\,\to\,p_7}%=\;e\,\frac{(2p_3+p_7)^{\mu}p_{7\,\mu}}{2p_3\cdot p_7}
\;=\;e\,\frac{2p_3\cdot p_7}{2p_3\cdot p_7}\;=\;e,\nn\\
s_{47}&=&e\,
\left.\frac{(-p_4-(p_4+p_7))^{\mu}\,\varepsilon_{\mu}^*(p_7)}{(p_4+p_7)^2-m_{\pi}^2}
\right|_{\varepsilon(p_7)\,\to\,p_7}%=\;-e\,\frac{(2p_4+p_7)^{\mu}p_{7\,\mu}}{2p_4\cdot p_7}
\;=\;-e\,\frac{2p_4\cdot p_7}{2p_4\cdot p_7}\;=\;-e.\nn
\eea
It is clear from the form of Eqs.~(\ref{a}) and (\ref{b}) that amplitudes
$M_a$ and $M_b$ neither vanish separately nor cancel each other, contrary
to the amplitudes of the other 20 Feynman diagrams of process (\ref{epmppmmg}) 
which also contain the vertices $\pi^{\mp}\gamma\rho^{\pm}$.
Although that degree of gauge invariance violation should not play any role
in practice, such effects should be treated with great care, as they may become
sizable in some regions of the photon phase space. Therefore, it is recommended
to use the {\tt igauge} option whenever new interaction vertices are added to
the program.

\begin{table}
\label{table}
\begin{center}
\begin{tabular}{c|cc|cc}
{\tt igauge }
& \multicolumn{2}{c|}{$\sigma(\epm\;\to\;\pi^+ \pi^- \pi^+ \pi^- \gamma)$}
& \multicolumn{2}{c}{$\sigma(\epm\;\to\;\pi^+ \pi^- \mu^+ \mu^- \gamma)$}\\[1mm]
\hline
\rule{0mm}{5mm}
0 & 11.86(5)      & 11.83(5)     & 0.0590(2)    & 0.0586(2)    \\
1 & 0.124(2)e-30  & 0.441(1)e-10 & 0.636(9)e-33 & 0.973(1)e-9 
\end{tabular}
\caption{\small The cross sections in pb of processes (\ref{epmppmmg}) 
and (\ref{epm4pg}) at $\sqrt{s}=1$~GeV without (first column) and with (second
column) contributions from the $\pi^{\mp}\gamma\rho^{\pm}$ interaction vertices
of Fig.~\ref{verts3}. The cuts used in the computation are given by 
(\ref{cuts}). The numbers in parentheses show the MC uncertainty of the last 
decimal.}
\end{center}
\end{table}

\section{Preparation for running and program usage}
%
%Other minor changes to the program, including corrections of a few 
%bugs are described in a {\tt readme} file. 

{\tt carlomat\_3.0} is distributed as a single {\tt tar.gz} archive
{\tt carlomat\_3.0.tgz} which can be downloaded from: 
http://kk.us.edu.pl/carlomat.html. When untared with a command\\
{\tt tar -xzvf carlomat\_3.0.tgz}\\
it will create directory {\tt carlomat\_3.0} with sub directories:
{\tt code\_generation}, {\tt mc\_computation}, {\tt carlolib},
{\tt test\_output} and {\tt test\_output0}.

Although {\tt carlomat\_3.0} is dedicated to the description of
low energy $\epm$ scattering, interfaces to the parton density functions
are kept. Therefore, files {\tt mstwpdf.f} of {\tt MSTW} \cite{MSTW}
and {\tt Ctq6Pdf.f} and {\tt cteq6l.tbl} of {\tt CTEQ6} \cite{CTEQ}
are also included in the current distribution of the program,
but grids must be downloaded from the web page of {\tt MSTW}, 
see {\tt readme} file or \cite{carlomat2} for details.
If the program will not be run for the hadron scattering processes the user
can comment lines contained between {\tt ckk\_had>} and {\tt ckk\_had<} 
in {\tt crosskk.f} and {\tt parfixkk.f}, and comment or remove references
to {\tt mstw\_interface.o, mstwpdf.o, Ctq6Pdf.o, ctq6f\_interface.o} from
{\tt makefile} in {\tt mc\_computation}.

Preparation for running requires basically the same steps as in 
{\tt carlomat\_2.0}. They are recollected below for user's convenience.
\begin{itemize}
\item Choose a Fortran 90 compiler in {\tt makefile}'s 
of {\tt code\_generation} and {\tt mc\_computation}
and compile all the 
routines of {\tt carlolib} with the same compiler as that chosen in 
{\tt mc\_computation}; 
\item Specify the process and required
options in {\tt carlomat.f} and execute 
{\tt make code} from the command line in {\tt code\_generation};
\item Go to {\tt mc\_computation}, choose the center of mass energy and
required options in {\tt carlocom.f}
and execute {\tt make mc} in the command line.
\end{itemize}
Whenever the Fortran compiler is changed, or a
compiled program is transferred to another computer with a different
processor, all the object and module files should be deleted 
by executing the commands:\\
{\tt rm *.o}\\
{\tt rm *.mod}\\
and the necessary steps of those listed above should be repeated.

The basic output of the MC run is written to file {\tt tot\_name}, where 
{\tt name} is created automatically if the assignment for character variable\\
{\tt prcsnm='auto'}\\
in {\tt carlomat.f} is not changed to arbitrary user's defined name.
The output files for processes (\ref{epmppmmg}) and (\ref{epm4pg}) with
the preselected parameters and options should reproduce those delivered in
directory {\tt test\_output0}. 

If the differential cross sections/distributions are required then set\\
{\tt idis=1} \\
in {\tt carlocom.f}. The number of distributions to be calculated
must be specified in {\tt distribs.f} and their parameters should be defined 
in {\tt calcdis.f}. The output will be stored in data files 
{\tt db\#\_name} and {\tt dl\#\_name} which can be plotted
with boxes and lines, respectively, with the use of {\tt gnuplot}. 
When the run is finished all output files, except for {\tt test} that may 
contain information relevant in case of unexpected program stop, are 
moved to directory {\tt test\_output}.

As in former versions of the program, there is a possibility of generating 
the unweighted events. It is governed by the option {\tt imc} that is
available in {\tt carlocom}.

The code generation for processes (\ref{epmppmmg}) and (\ref{epm4pg})
takes a fraction of a second time.
The MC computation of the cross sections of Table~\ref{table}
in 10 iterations, with a maximum of $200\,000$ calls to the integrand each,
takes 142s and 43s time, respectively, for (\ref{epmppmmg}) and (\ref{epm4pg}) 
on processor Intel$^{\tt R}$ Core$^{\tt TM}$ i5-4200M CPU 
@ 2.50~GHz with a 64 bit Intel Fortran compiler.

{\bf Acknowledgement:} This project was supported in part with financial 
resources of the Polish National Science Centre (NCN) under grant decision 
No. DEC-2011/03/B/ST6/01615. The author is grateful to Fred Jegerlehner for
providing the Feynman rules of the HLS
model and to Henryk Czy\.z and Szymon Tracz for providing the results of 
fits of the nucleon form factors.

\end{document}